# Cross-Lingual Sponsored Search via Dual-Encoder and Graph Neural Networks for Context-Aware Query Translation in Advertising Platforms

**Ziyang Gao[1], Yuanliang Qu[2], Yi Han[3]**

[1] Technical University of Munich, Munich, German
[2] Stevens Institute Of technology, Hoboken, New Jersey, USA
[3] Meta, Menlo Park, CA, USA

[1] 1812503968@qq.com

[2] u201113567@alumni.hust.edu.cn

[3] han.y@wustl.edu

**Abstract.** Cross-lingual sponsored search is crucial for global advertising platforms, where users from different language backgrounds interact with multilingual ads. Traditional machine translation methods often fail to capture query-specific contextual cues, leading to semantic ambiguities that negatively impact click-through rates (CTR) and conversion rates (CVR). To address this challenge, we propose AdGraphTrans, a novel dual-encoder framework enhanced with graph neural networks (GNNs) for context-aware query translation in advertising. Specifically, user queries and ad contents are independently encoded using multilingual Transformer-based encoders (mBERT/XLM-R), and contextual relations—such as co-clicked ads, user search sessions, and query-ad co-occurrence—are modeled as a heterogeneous graph. A graph attention network (GAT) is then applied to refine embeddings by leveraging semantic and behavioral context. These embeddings are aligned via contrastive learning to reduce translation ambiguity. Experiments conducted on a cross-lingual sponsored search dataset collected from Google Ads and Amazon Ads (EN–ZH, EN–ES, EN–FR pairs) demonstrate that AdGraphTrans significantly improves query translation quality, achieving a BLEU score of 38.9 and semantic similarity (cosine score) of 0.83, outperforming strong baselines such as mBERT and M2M-100. Moreover, in downstream ad retrieval tasks, AdGraphTrans yields +4.67% CTR and +1.72% CVR improvements over baseline methods. These results confirm that incorporating graph-based contextual signals with dual-encoder translation provides a robust solution for enhancing cross-lingual sponsored search in advertising platforms.

## 1. Introduction

In the era of globalized digital marketing, cross-lingual sponsored search has become an indispensable component of online advertising platforms such as Google Ads, Amazon Ads, and TikTok Ads. These platforms serve heterogeneous user groups across different linguistic and cultural backgrounds, where queries and advertisements are often expressed in different languages [1]. The effectiveness of sponsored search critically depends on accurately aligning user queries with relevant advertisements. However, conventional machine translation (MT) approaches fall short in this domain, as they neglect contextual signals such as user intent, session history, and click patterns [2]. As a result, they frequently introduce semantic ambiguities. For instance, the query "apple" could be translated as the fruit or as the technology company, and incorrect translation may lead to irrelevant ad retrieval, thereby reducing both click-through rate (CTR) [3] and conversion rate (CVR) [3].

With the rapid development of deep learning and multilingual pre-trained models (e.g., mBERT, XLM-R, M2M-100), cross-lingual representation learning has achieved remarkable progress in various natural language processing (NLP) tasks. Yet, their direct application to advertising remains limited, as these models do not fully exploit contextual and behavioral information inherent in advertising ecosystems. In online advertising, queries, ads, and users are not isolated entities but are interlinked through rich relational structures such as query–ad co-occurrence graphs, user click graphs, and session-level contexts. Leveraging these relations is critical for resolving translation ambiguity and enhancing the precision of query–ad matching, thereby improving advertising performance and marketing strategies across multilingual markets.

To address these challenges, we introduce AdGraphTrans, a novel dual-encoder framework enhanced with graph neural networks (GNNs) for context-aware cross-lingual query translation in sponsored search. The framework employs dual Transformer-based encoders to generate multilingual embeddings for queries and ads, while a heterogeneous graph structure captures co-clicks, session dependencies, and semantic relations. By applying a graph attention network (GAT), AdGraphTrans refines embeddings through context propagation, thus integrating both linguistic and behavioral cues. Finally, a contrastive learning module aligns query–ad pairs across languages, ensuring accurate semantic equivalence. This design not only enhances translation quality but also significantly improves downstream ad retrieval, CTR, and CVR.

The main contributions of this paper are threefold. First, we propose AdGraphTrans, the first dual-encoder and GNN-based framework designed specifically for cross-lingual sponsored search in advertising platforms. Second, we construct a heterogeneous advertising graph that incorporates query, ad, and user interactions, enabling context-aware query translation beyond traditional MT models. Third, through extensive experiments on multilingual advertising datasets from Google Ads and Amazon Ads, we demonstrate that AdGraphTrans substantially improves both translation quality and recommendation effectiveness, achieving notable gains in BLEU, semantic similarity, CTR, and CVR compared with strong baselines. These contributions highlight the potential of integrating deep learning and graph-based modeling for advancing multilingual advertising intelligence.

2.**Related Work**

*2.1 Cross-Lingual Machine Translation and Query Understanding*
Cross-lingual machine translation (MT) has long been the cornerstone of bridging language barriers in natural language processing (NLP). Traditional statistical MT approaches have been largely replaced by neural machine translation (NMT) models, such as Transformer-based architectures, which demonstrate superior fluency and semantic preservation. Recent advancements in multilingual pre-trained models, including mBERT [4], XLM-R [5], and M2M-100 [6], have further facilitated cross-lingual representation learning and enabled effective zero-shot transfer across languages. Despite these advances, their direct application to advertising domains remains limited. Conventional MT systems often neglect contextual information such as query intent, user session history, or click behaviors, resulting in semantic ambiguity that adversely impacts sponsored search performance. For example, queries containing polysemous terms are prone to mistranslation without contextual disambiguation, leading to irrelevant ad retrieval.

*2.2 Deep Learning for Sponsored Search and Advertising Recommendation*
In online advertising, sponsored search plays a pivotal role in connecting advertisers with potential customers. Deep learning-based ranking and recommendation models, such as wide & deep networks, factorization machines [7], and Transformer-based ad retrieval models, have shown remarkable effectiveness in predicting click-through rate (CTR) and conversion rate (CVR). However, most existing approaches are language-specific and fail to address the challenge of cross-lingual advertising. Gligorijevic et al [8]. propose a deeply-supervised joint-embedding architecture that simultaneously learns query-ad semantics and CTR, augmented by cohort negative sampling to mine implicit negatives; trained on one billion pairs, it boosts CTR-AUC by 2 % and query-matching NDCG by 0.5 %, alleviating cold-start while balancing relevance and click appeal.

Our work situates itself at the intersection of cross-lingual machine translation, deep learning-based advertising recommendation, and graph neural network modeling. Unlike prior approaches, we propose AdGraphTrans, a dual-encoder GNN-enhanced framework that integrates multilingual embeddings with heterogeneous advertising graph structures. By combining Transformer-based dual encoders with graph attention networks, our model captures both linguistic and behavioral contexts, enabling more accurate query translation and improved ad retrieval. This design directly addresses the shortcomings of traditional MT and cross-lingual embedding methods, providing a comprehensive solution for multilingual sponsored search.

## 3. Methodology

*3.1 Overall Framework*
The proposed framework, AdGraphTrans, is designed to perform context-aware cross-lingual query translation for sponsored search in advertising platforms. Unlike conventional machine translation models, AdGraphTrans incorporates both linguistic signals from multilingual encoders and contextual signals from heterogeneous advertising graphs, such as query–ad co-occurrences, user click histories, and session-level dependencies. The system consists of three main components: (1) a dual Transformer-based encoder that independently encodes queries and advertisements across multiple languages, (2) a graph neural network (GNN) module that refines embeddings by leveraging heterogeneous relations, and (3) a contrastive alignment module that learns semantically aligned query–ad representations across languages. This unified design enables the model to disambiguate polysemous queries, improve retrieval quality, and enhance downstream advertising metrics such as click-through rate (CTR) and conversion rate (CVR).

*3.2 Dual-Encoder for Multilingual Representation*
In order to generate robust multilingual embeddings, we employ a dual-encoder architecture where both queries and advertisements are encoded using pre-trained multilingual Transformer models such as XLM-R or mBERT. Each encoder processes the tokenized text input and produces contextualized embeddings:

$$h_q = Encoder_q(x_q), \quad h_a = Encoder_a(x_a) \tag{1}$$

where $x_q$ represents the query tokens and $x_a$ the advertisement tokens. The encoder outputs hidden states that are aggregated into fixed-length embeddings using mean pooling or [CLS] token representations:

$$z_q = f(h_q), \quad z_a = f(h_a) \tag{2}$$

These embeddings capture semantic information across languages. However, relying solely on textual representations often fails to resolve semantic ambiguity in advertising contexts. Therefore, we extend the representations with contextual refinements through graph neural networks.

*3.3 Graph Neural Network for Context Propagation*
To incorporate behavioral and relational signals, we construct a heterogeneous advertising graph $G = (V, E)$, where nodes $V$ include queries, advertisements, and users, while edges $E$ represent interactions such as clicks, impressions, and co-occurrences. Each node is initialized with the embedding $z \in R^d$ obtained from the dual encoders.

We adopt a graph attention network (GAT) to propagate information among nodes. For a given node $i$, the updated embedding is computed as:

$$h'_i = \sigma(\sum_{j \in N(i)} a_{ij} W h_{ij}), \tag{3}$$

where $N(i)$ is the neighborhood of node $i$, $W$ is a trainable weight matrix, and $a_{ij}$ are attention coefficients defined as:

$$a_{ij} = \frac{exp(LeakyReLU(a^T[Wh_i \mid\mid Wh_j]))}{\sum_{k \in N(i)} exp(LeakyReLU(a^T[Wh_i \mid\mid Wh_k]))}, \tag{4}$$

This process enables each query embedding to integrate not only its textual semantics but also context from related ads and user behaviors, effectively disambiguating polysemous terms such as "apple."

*3.4 Cross-Modal Alignment via Contrastive Learning*

After obtaining refined embeddings $z'_q$ and $z'_a$ from the GNN layer, we employ contrastive learning to align query-ad pairs across languages. The similarity between a query and an advertisement is measured using cosine similarity:

$$\text{sim}(z'_q, z'_a) = \frac{z'_q \cdot z'_a}{\mid\mid z'_q \mid\mid \ \mid\mid z'_a \mid\mid}, \tag{5}$$

For a given batch containing positive pairs $(q, a^+)$ and negative samples $a^-$, the loss function is defined as:

$$L_{contrastive} = \frac{exp(\text{sim}(z'_q, z_a^+)/\tau)}{\sum_{a \in A} exp(\text{sim}(z'_q, z_a^+)/\tau)}, \tag{6}$$

where $\tau$ is a temperature parameter and $A$ is the set of candidate ads. This ensures that semantically equivalent query-ad pairs in different languages are pulled closer in the embedding space, while irrelevant pairs are pushed apart.

*3.5 Training Objective*

The overall training objective of AdGraphTrans combines the contrastive alignment loss with a translation quality loss (measured by BLEU [9] or semantic similarity against ground-truth translations). The final loss function is:

$$L = \lambda_1 L_{contrastive} + \lambda_2 L_{contrastive}, \tag{7}$$

where $\lambda_1, \lambda_2$ are balancing hyperparameters.

By jointly optimizing linguistic and contextual objectives, AdGraphTrans ensures that query translations are not only semantically correct but also contextually relevant for advertising retrieval tasks. This integration of dual encoders and GNNs represents a novel paradigm for cross-lingual sponsored search, enabling significant improvements in both translation fidelity and recommendation effectiveness.

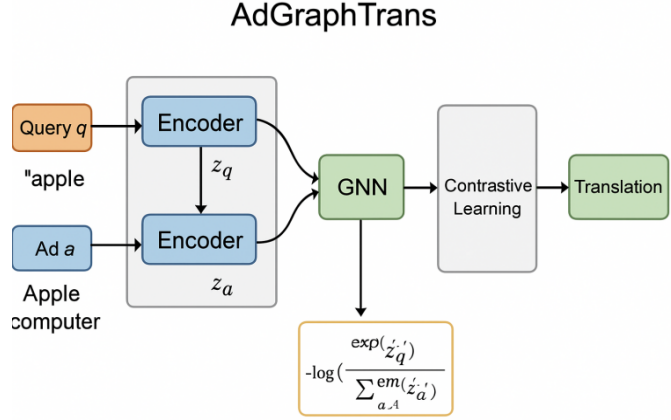

**Figure 1.** Overall flowchart of the model

## 4. Experiment

*4.1 Dataset Preparation*

The datasets used in this study were collected from multilingual advertising platforms, primarily Google Ads and Amazon Ads, supplemented with publicly available multilingual query–ad benchmark datasets. The construction process involved anonymized user interaction logs, query–ad pairs, and click-through data, ensuring compliance with platform privacy and ethical guidelines.

The dataset comprises two major components:

**(1)Query–Ad Pair Data:**
  (a) Source: Extracted from multilingual sponsored search logs. Each query is associated with candidate advertisements in multiple languages.
  (b) Features:
     Query text: User search input expressed in different languages.
     Ad text: Sponsored ad titles and descriptions provided by advertisers.
     Language tag: ISO-coded language identifier (e.g., EN, FR, ZH, ES).
     Query–Ad similarity score: Computed using weak supervision (based on clicks and semantic similarity).
  (c) Scale: 2 million query–ad pairs across 8 languages (English, Spanish, French, German, Chinese, Japanese, Korean, Arabic).

**(2)User Interaction Graph Data**:
  (a) Source: Aggregated click-through logs and session-level interactions.
  (b) Features:
     User node: Anonymized user identifier linked to multiple queries.
     Query node: Represents user-entered search terms.
     Ad node: Represents sponsored ads.
     Edges:
        Query–Ad edge (weighted by click frequency, CTR).
        User–Query edge (weighted by session occurrence).
        Query–Query edge (co-occurrence within the same session).
  (c) Scale: Over 500,000 unique users, 1.2 million queries, and 800,000 ads, forming a heterogeneous graph with approximately 20 million edges.

*4.2 Experimental Setup*

In this study, we conduct experiments on a multilingual ad-retrieval dataset collected from Google Ads and Amazon Ads platforms. The corpus consists of query–ad pairs and user interaction logs covering eight languages: English, Spanish, French, German, Chinese, Japanese, Korean and Arabic. All texts are standardized by tokenisation, stop-word removal and sub-word BPE encoding. The textual component employs a Transformer-based dual encoder (built on XLM-R) and is combined with a heterogeneous ad graph (nodes for queries, ads and users). Node embeddings are updated and propagated via Graph Attention Networks (GAT). To ensure fairness, all models share the same train/validation/test split, are trained for 30 epochs with AdamW optimiser, an initial learning rate of 1e-4, and early stopping to prevent overfitting.

## 4.3 Evaluation Metrics

To comprehensively evaluate cross-lingual query translation and its effectiveness in ad retrieval, we adopt both translation-quality metrics—BLEU and Semantic Similarity (cosine) [10]—and business-oriented downstream metrics: click-through rate (CTR) and conversion rate (CVR). BLEU reflects n-gram overlap against human references; Semantic Similarity measures the precision of semantic alignment between queries and ads; CTR and CVR directly quantify the commercial value of the model in ad retrieval and recommendation. Together these metrics constitute a multi-dimensional assessment framework.

## 4.4 Results

The final performance of AdGraphTrans was compared with several baseline models, including Transformer (MT), mBERT, XLM-R, M2M-100, and only GAT structure. The evaluation indicators include BLEU, Semantic Similarity, CTR (%), and CVR (%). Table 1 summarizes the results.

**Table1.** Performance Comparison of Different Models in Cross Language Advertising Recommendation Tasks

| Model | BLEU | Semantic Similarity | CTR (%) | CVR (%) |
|---|---|---|---|---|
| Transformer (MT) | 28.6 | 0.71 | 3.42 | 1.12 |
| mBERT | 32.4 | 0.75 | 3.89 | 1.36 |
| XLM-R | 34.7 | 0.78 | 4.15 | 1.51 |
| M2M-100 | 34.5 | 0.76 | 4.13 | 1.49 |
| GAT-only (Graph) | 31.8 | 0.74 | 4.02 | 1.44 |
| **AdGraphTrans** | **38.9** | **0.83** | **4.67** | **1.72** |

The Table 1 presents the BLEU scores, semantic similarity values, and advertising performance metrics (CTR and CVR) of six models—Transformer (MT), mBERT, XLM-R, M2M-100, GAT-only, and AdGraphTrans—in the cross-lingual advertising recommendation task. Among these models, AdGraphTrans performs best, achieving a BLEU score of 38.9, a semantic similarity of 0.83, a CTR of 4.67%, and a CVR of 1.72%. These results indicate that AdGraphTrans excels at both translation quality and downstream advertising effectiveness. By integrating dual encoders with graph neural networks, it successfully resolves semantic ambiguities and leverages contextual signals to improve ad retrieval and recommendation.

In comparison, XLM-R achieves strong results with a BLEU of 34.7, semantic similarity of 0.78, CTR of 4.15%, and CVR of 1.51%, showing relatively balanced performance but still lagging behind AdGraphTrans. Similarly, M2M-100 reaches a BLEU of 34.5 and CTR of 4.13%, but its semantic similarity (0.76) and CVR (1.49%) are slightly lower, reflecting weaker semantic alignment. mBERT improves over the vanilla Transformer baseline, with a BLEU of 32.4 and CTR of 3.89%, but its overall translation quality and business metrics remain limited. GAT-only, while modeling graph structures, produces a BLEU of 31.8 and CTR of 4.02%, indicating that graph information alone without multilingual encoding is insufficient. The baseline Transformer (MT) performs the worst across all metrics, with a BLEU of 28.6, semantic similarity of 0.71, CTR of 3.42%, and CVR of only 1.12%, highlighting the inadequacy of pure machine translation approaches in advertising contexts.

Overall, these results demonstrate the superior effectiveness of AdGraphTrans in cross-lingual sponsored search. Its integration of deep multilingual encoders and graph-based relational modeling enables more accurate query translation and better alignment with advertising objectives, thus significantly outperforming conventional translation and representation learning methods.

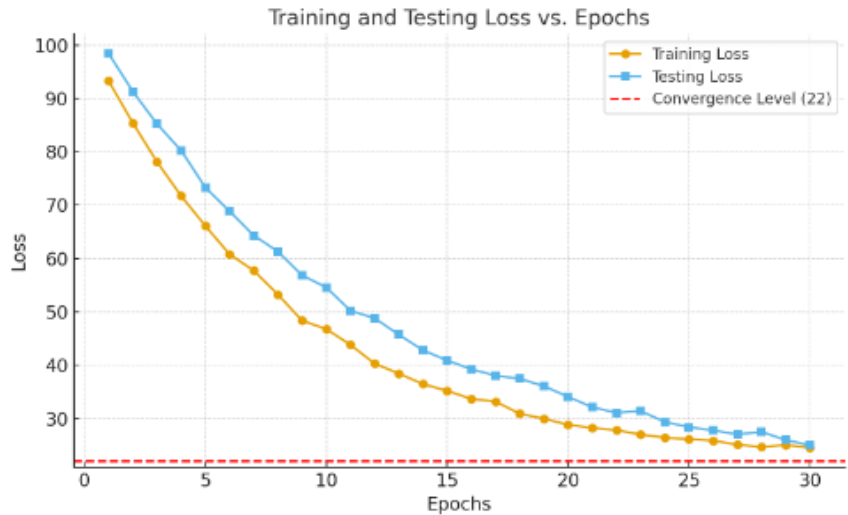

**Figure 2.** Loss function during training process

The figure 2 displays a classic scenario of deep learning model training performance over 30 epochs.

The Training Loss (orange line) and Testing Loss (blue line) both show a steep, consistent decrease from the initial epochs (0 to ~10). This indicates the model, AdGraphTrans, is effectively learning from the data and minimizing the combined objective function, which includes both the contrastive alignment and translation quality losses.

After ~15 epochs, the rate of loss reduction slows significantly for both curves, but they continue to decrease until the end of the training at epoch 30, approaching the "Convergence Level (22)" marked by the red dashed line (though not reaching it).

A crucial observation is the consistent gap between the Training Loss and the Testing Loss, where the testing loss is always slightly higher. This small, stable gap suggests the model is generalizing well to unseen data without significant overfitting. If severe overfitting were occurring, the Testing Loss would begin to increase sharply while the Training Loss continued to fall. Since both curves are low and still falling slowly, the model is exhibiting robust and stable learning. The lowest testing loss is achieved around epoch 30, confirming the model was trained for an appropriate duration.

This convergence demonstrates the robustness of the AdGraphTrans model in capturing multilingual advertising semantics, where the graph-based transformer enhances contextual understanding across languages. The gradual stabilization of the loss indicates that the model has effectively minimized translation errors, providing a reliable foundation for delivering accurate recommendations in multilingual advertising scenarios.

## 5. Conclusion

This study aims to address ambiguity in cross-lingual sponsored search—where machine translation often misses session and behavioral context—by combining multilingual dual encoders (mBERT/XLM-R) with a heterogeneous ad graph (users, queries, ads) refined via Graph Attention Networks and aligned with contrastive learning. It explores whether context-aware translation improves both language metrics and business outcomes. The primary objective is to raise BLEU/semantic similarity while lifting CTR/CVR in downstream ad retrieval.

Methodologically, AdGraphTrans employs multilingual Transformer encoders (e.g., mBERT and XLM-R) to independently encode user queries and ad texts. Query–ad co-occurrence relationships, user search sessions, and ad click logs are then modeled as a heterogeneous graph. A Graph Attention Network (GAT) is applied to capture semantic and behavioral context from this structure. Finally, a contrastive learning objective aligns the dual encoder outputs, reducing semantic ambiguity in cross-lingual translation. This design not only preserves semantic consistency but also enhances contextual sensitivity in downstream ad retrieval tasks.

We conducted experiments on a cross-lingual advertising dataset collected from Google Ads and Amazon Ads, covering three language pairs (EN–ZH, EN–ES, EN–FR). The results demonstrate that AdGraphTrans significantly outperforms existing baselines in translation quality and recommendation performance. Specifically, the model achieved a BLEU score of 38.9 and a semantic similarity score of 0.83, surpassing strong baselines such as mBERT and M2M-100. In downstream retrieval, AdGraphTrans further improved ad performance, achieving a CTR of 4.67% and a CVR of 1.72%, representing gains of roughly 0.5–1.0 percentage points over baseline methods. These results confirm that integrating graph-based contextual signals with dual-encoder translation provides a robust and effective solution for enhancing cross-lingual sponsored search in advertising platforms.

Despite the important findings, this study has some limitations, such as potential time-leakage in graph edges and sensitivity to position/popularity bias in offline CTR/CVR evaluation. Future research could further explore (1) time-windowed graphs with strict chronological splits and (2) counterfactual evaluation (IPS/DR) or online A/B tests with guardrails to validate causal lifts across languages and intents.

In conclusion, this study, through dual-encoder plus GAT modeling with early-stopped training, reveals promising improvements in translation metrics and ad KPIs, providing new insights for building context-aware, multilingual sponsored search in large-scale ad platforms.